\documentclass[aps,prx,twocolumn,footinbib,superscriptaddress,floatfix]{revtex4-2}

\usepackage{CJK}
\usepackage[colorlinks,bookmarks=false,citecolor=blue,linkcolor=red,urlcolor=blue]{hyperref}
\usepackage{color} 
\usepackage{graphicx}
\usepackage{float}
\usepackage{multirow}
\usepackage{amsmath}
\usepackage{bm}
\usepackage{amsfonts}
\usepackage{subfigure}
\usepackage{cleveref}

\usepackage{dcolumn}
\usepackage{amssymb}
\usepackage{microtype}
\usepackage{xfrac}
\usepackage{array}
\usepackage[dvipsnames]{xcolor}

\newcolumntype{P}[1]{>{\centering\arraybackslash}p{#1}}

\begin{document}

\title{Reply to ``Comment on: `Case for a U(1)$_\pi$ Quantum Spin Liquid Ground State \\ in the Dipole-Octupole Pyrochlore Ce$_2$Zr$_2$O$_7$' "}

\author{E.~M.~Smith}
\affiliation{Department of Physics and Astronomy, McMaster University, Hamilton, Ontario L8S 4M1, Canada}
\affiliation{Brockhouse Institute for Materials Research, McMaster University, Hamilton, Ontario L8S 4M1, Canada}

\author{O.~Benton}
\affiliation{Max Planck Institute for the Physics of Complex Systems, N\"{o}thnitzer Stra{\ss}e 38, Dresden 01187, Germany}

\author{D.~R.~Yahne}
\affiliation{Department of Physics, Colorado State University, 200 West Lake Street, Fort Collins, Colorado 80523-1875, USA}

\author{B.~Placke}
\affiliation{Max Planck Institute for the Physics of Complex Systems, N\"{o}thnitzer Stra{\ss}e 38, Dresden 01187, Germany}

\author{R.~Sch\"{a}fer}
\affiliation{Max Planck Institute for the Physics of Complex Systems, N\"{o}thnitzer Stra{\ss}e 38, Dresden 01187, Germany}

\author{J.~Gaudet}
\affiliation{Department of Physics and Astronomy, McMaster University, Hamilton, Ontario L8S 4M1, Canada}

\author{J.~Dudemaine}
\affiliation{D\'epartement de Physique, Universit\'e de Montr\'eal, Montr\'eal, Canada}
\affiliation{Regroupement Qu\'eb\'ecois sur les Mat\'eriaux de Pointe (RQMP)}

\author{A.~Fitterman}
\affiliation{D\'epartement de Physique, Universit\'e de Montr\'eal, Montr\'eal, Canada}
\affiliation{Regroupement Qu\'eb\'ecois sur les Mat\'eriaux de Pointe (RQMP)}

\author{J.~Beare}
\affiliation{Department of Physics and Astronomy, McMaster University, Hamilton, Ontario L8S 4M1, Canada}

\author{A.~R.~Wildes}
\affiliation{Institut Laue-Langevin, 71 Avenue des Martyrs CS 20156, 38042 Grenoble Cedex 9, France}

\author{S.~Bhattacharya}
\affiliation{Universit\'e Paris-Saclay, CNRS, Laboratoire de Physique des Solides, 91405 Orsay, France}

\author{T.~DeLazzer}
\affiliation{Department of Physics, Colorado State University, 200 West Lake Street, Fort Collins, Colorado 80523-1875, USA}

\author{C.~R.~C.~Buhariwalla}
\affiliation{Department of Physics and Astronomy, McMaster University, Hamilton, Ontario L8S 4M1, Canada}

\author{N.~P.~Butch}
\affiliation{Center for Neutron Research, National Institute of Standards and Technology, MS 6100 Gaithersburg, Maryland 20899, USA}

\author{R.~Movshovich}
\affiliation{Los Alamos National Laboratory, Los Alamos, New Mexico 87545, USA}

\author{J.~D.~Garrett}
\affiliation{Brockhouse Institute for Materials Research, McMaster University, Hamilton, Ontario L8S 4M1, Canada}

\author{C.~A.~Marjerrison}
\affiliation{Brockhouse Institute for Materials Research, McMaster University, Hamilton, Ontario L8S 4M1, Canada}

\author{J.~P.~Clancy}
\affiliation{Department of Physics and Astronomy, McMaster University, Hamilton, Ontario L8S 4M1, Canada}
\affiliation{Brockhouse Institute for Materials Research, McMaster University, Hamilton, Ontario L8S 4M1, Canada}

\author{E.~Kermarrec}
\affiliation{Universit\'e Paris-Saclay, CNRS, Laboratoire de Physique des Solides, 91405 Orsay, France}

\author{G.~M.~Luke}
\affiliation{Department of Physics and Astronomy, McMaster University, Hamilton, Ontario L8S 4M1, Canada}
\affiliation{Brockhouse Institute for Materials Research, McMaster University, Hamilton, Ontario L8S 4M1, Canada}

\author{A.~D.~Bianchi}
\affiliation{D\'epartement de Physique, Universit\'e de Montr\'eal, Montr\'eal, Canada}
\affiliation{Regroupement Qu\'eb\'ecois sur les Mat\'eriaux de Pointe (RQMP)}

\author{K.~A.~Ross}
\affiliation{Department of Physics, Colorado State University, 200 West Lake Street, Fort Collins, Colorado 80523-1875, USA}
\affiliation{Canadian Institute for Advanced Research, 661 University Avenue, Toronto, Ontario M5G 1M1, Canada.}

\author{B.~D.~Gaulin}
\affiliation{Department of Physics and Astronomy, McMaster University, Hamilton, Ontario L8S 4M1, Canada}
\affiliation{Brockhouse Institute for Materials Research, McMaster University, Hamilton, Ontario L8S 4M1, Canada}
\affiliation{Canadian Institute for Advanced Research, 661 University Avenue, Toronto, Ontario M5G 1M1, Canada.}

\date{\today}

\begin{abstract} 
In his comment [\href{https://arxiv.org/abs/2209.03235}{arXiv:2209:03235}], S. W. Lovesey argues that our analysis of neutron scattering 
experiments performed on Ce$_2$Zr$_2$O$_7$ is invalid.
Lovesey argues that we have not properly accounted for the higher-order multipolar contributions to the magnetic scattering
and that our use of pseudospin-$1/2$ operators to describe the scattering is inappropriate.
In this reply, we show that the multipolar corrections discussed by Lovesey only become significant at scattering wavevectors exceeding those accessed in our experiments.
This in no way contradicts or undermines our work, which never claimed a direct observation of scattering from higher-order
multipoles. 
We further show that Lovesey's objections to our use of pseudospins are unfounded, and that the pseudospin operators
are able to describe all magnetic scattering processes at the energy scale of our experiments, far below
the crystal field gap. 
Finally, we comment on certain assumptions in Lovesey's calculations of the scattering amplitude which are inconsistent
with experiment.
\end{abstract}
\maketitle

The comment [\onlinecite{lovesey-comment}] presents two principal objections to our original study 
 [\onlinecite{original-paper}]:
 \begin{enumerate}
 \item{It is argued that we did not properly account for corrections to the neutron scattering cross section coming from higher-order multipoles (beyond the dipolar approximation).}
 \item{It is argued that using pseudospin degrees of freedom to describe magnetic scattering is fundamentally inappropriate, since the pseudospins {\it ``are irrelevant in the material world''} \cite{lovesey-comment}.}
 \end{enumerate}
 
\begin{figure}
\centering
\includegraphics[width=\columnwidth]{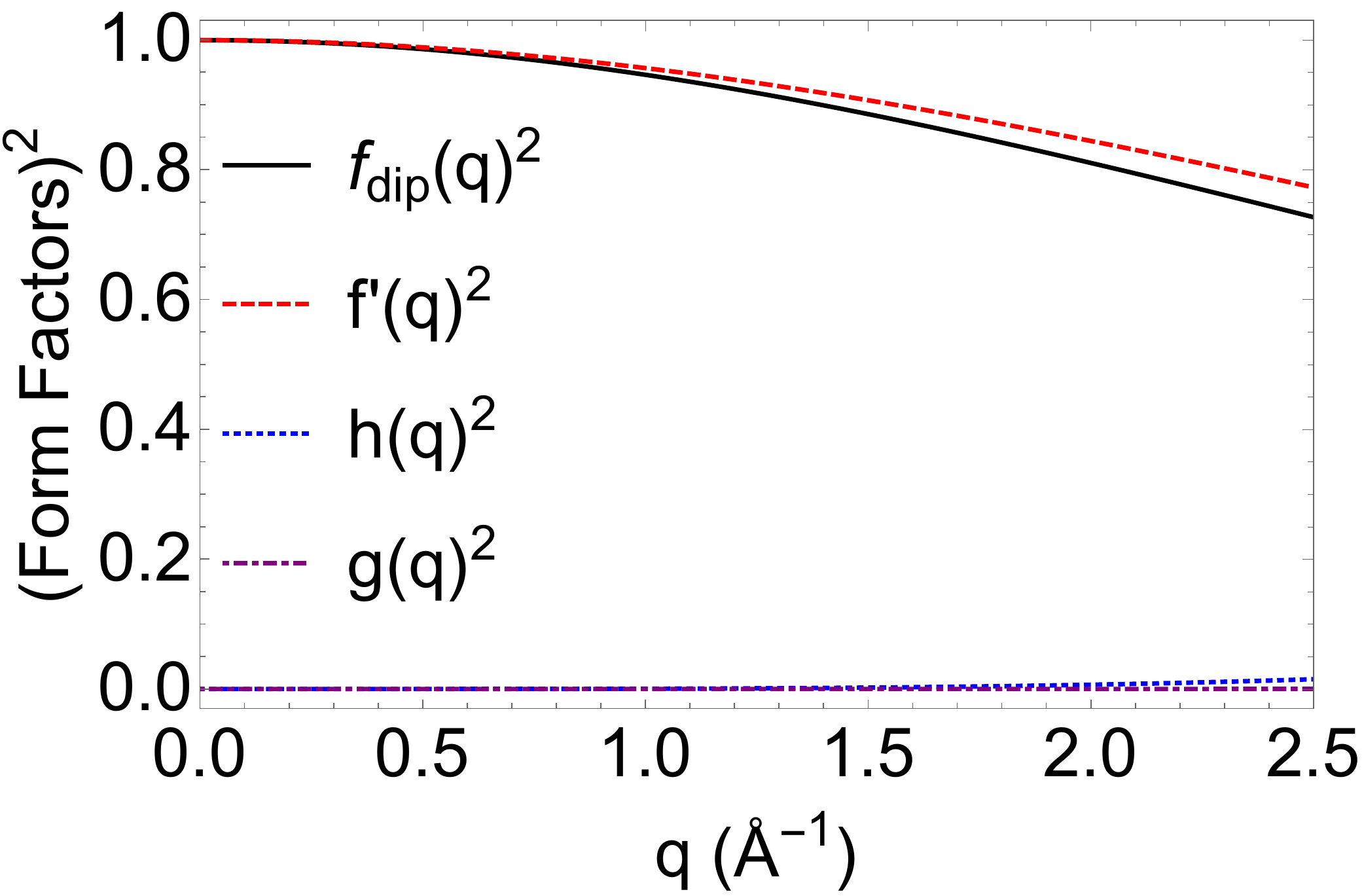}
\caption{Squares of form factors multiplying different multipolar contributions to the scattering amplitude for magnetic scattering from a
Ce$^{3+}$ ion, over the momentum range $q \in [0, 2.5] {\rm \AA}^{-1}$ relevant to the experiments in [\onlinecite{original-paper}].
The form factor arising from the dipolar approximation $f_{\rm dip}(q)$ [Eq. (\ref{eq:Mbasic})], used in our calculation, agrees closely with $f'(q)$, the form factor of the leading, dipolar, contribution in Eq. (\ref{eq:M_complicated}). The form factors of the higher order
multipoles $h(q)$ and $g(q)$ are both small over this momentum range.
This justifies the use of the dipolar approximation in theoretical calculations of the scattering, and demonstrates that the higher order multipole corrections discussed in [\onlinecite{lovesey-comment}] will become significant only at larger wavevectors.
Calculations are based on analytical approximations to the functions $\langle j_n(q) \rangle$ given in [\onlinecite{Lisher1971}].
}
\label{fig:formfactors}
\end{figure}

 In this response, we will show that:
  \begin{enumerate}
 \item{Higher order multipolar corrections to the scattering intensity are insignificant over the range of scattering wavevectors probed in
 our experiments. 
 This can be seen by calculating the
 form factors for the higher order
 contributions [see Fig. \ref{fig:formfactors}].
 Directly observing the higher multipolar scattering was not an aim of our study, 
 and we did not claim any such observation.
 As such, our analysis of the scattering, in which we work within the dipole approximation, is appropriate.}
 \item{The use of pseudospin-$1/2$ operators, which is standard in the literature on rare-earth frustrated magnetism  \cite{Ross2011, Savary2012_er, Benton2016, Li2015, rau16, Wen2017, RauReview2019, sarkis20, Sibille2020, Dun2020, samartzis22} , 
 is the appropriate way to treat the low energy magnetic degrees of freedom in Ce$_2$Zr$_2$O$_7$. Far from being {\it ``irrelevant''}, these
 operators are directly related to measurable quantities and can account for all magnetic scattering processes (including multipolar scattering) which involve only the states of the low energy Kramers doublet (which is all of the magnetic scattering presented in \cite{original-paper}).}
 \item{The calculation of the scattering amplitude in [\onlinecite{lovesey-comment}] effectively assumes time-reversal symmetry breaking order, which is inconsistent with the experimental results.}
 \end{enumerate}
 
\section{Goals and outcomes of our study}
\label{sec:aims}

It appears to us that the comment [\onlinecite{lovesey-comment}]  is partly motivated  by a misunderstanding of the claims in [\onlinecite{original-paper}].
We therefore begin the response by briefly reviewing what we actually set out to do, and achieved, in [\onlinecite{original-paper}]. 
This will then set the context for the remainder of our response.

Ce$_2$Zr$_2$O$_7$ was identified in previous experimental studies as a highly frustrated magnet and quantum spin liquid (QSL) candidate \cite{Gao2019, Gaudet2019}. 
Neutron scattering showed that the crystal electric fields (CEF) on the Ce$^{3+}$ sites separate a low energy Kramers doublet from the rest of the CEF spectrum, by a large gap $\Delta_{\rm CEF} \approx 56 {\rm meV}$ \cite{Gao2019, Gaudet2019}.
Characterization of the wave-functions of the low energy CEF states revealed that they transform according to the $\Gamma_5^{+}$ and $\Gamma_6^{+}$ irreducible representations of the $D_{3d}$ point group.
This characterisation identifies Ce$_2$Zr$_2$O$_7$  as belonging to the family of dipolar-octupolar pyrochlores, whose novel properties were originally predicted in [\onlinecite{Huang2014}].

In particular, the symmetry allowed nearest-neighbor exchange interactions in such a material lead to a Hamiltonian [Eq. (1) of [\onlinecite{original-paper}]]
which has been predicted to give rise to a series of QSL states, occupying a large proportion of parameter space \cite{Benton2020, Patri2020}.
Given this context, the aims of our study were:
\begin{enumerate}
\item{To further characterize the low temperature state of Ce$_2$Zr$_2$O$_7$ using heat capacity, susceptibility and neutron
scattering measurements}
\item{To use these measurements to constrain the exchange parameters entering into the Hamiltonian describing Ce$_2$Zr$_2$O$_7$.}
\item{To place those estimates within published phase diagrams of the Hamiltonian, and thereby arrive at an inference of the
ground state.}
\end{enumerate}
These aims were achieved, resulting in the prediction that the ground state of Ce$_2$Zr$_2$O$_7$ is a $U(1)_{\pi}$ spin liquid.

We emphasise that, contrary to the depiction in [\onlinecite{lovesey-comment}], it was not our goal to provide further evidence that
Ce$_2$Zr$_2$O$_7$ is a dipolar-octupolar pyrochlore. We regard this as having already been proven by the experiments measuring
the CEF states \cite{Gao2019, Gaudet2019}, and we simply assume it for the purpose of our study.

Further, again contrary to [\onlinecite{lovesey-comment}], it was also not our goal to directly 
measure the higher order multipolar contribution to the neutron scattering intensity (unlike e.g. [\onlinecite{Sibille2020}]) and we did not claim to do so.
 In order to do this, we would have needed to measure scattering at much larger scattering 
wavevectors   (see Section \ref{sec:multipoles}). 
Measuring such scattering was not necessary for the purposes for which we used 
the neutron scattering data in our study, as discussed below.

We now turn to discuss the role that the theoretical calculations of the neutron scattering response played in our study.
The initial estimation of the exchange parameters was actually made by fitting heat capacity and susceptibility measurements
to model calculations, not by using the scattering data (again, contrary to the description of our work in [\onlinecite{lovesey-comment}]).
It was only once optimal sets of parameters had already been determined, that we compared model calculations of the neutron
scattering intensity to measurements.

This comparison served two purposes: to differentiate between two local optima in the goodness-of-fit with respect to the thermodynamic measurements and as an additional check on the consistency of our theory.
For the first purpose, it was sufficient to use data at relatively small momentum transfer (in the range $[0.43, 0.96] {\rm \AA}^{-1}$, see Fig. 10 of 
[\onlinecite{original-paper}]).
For the second purpose, we used all the data at hand, and this involved a range of momenta $q \lesssim 2.5 {\rm \AA}^{-1}$.
Although at such wavevectors the scattering is only meaningfully sensitive to magnetic dipoles (see Section \ref{sec:multipoles})
this is nevertheless a nontrivial test of the theory as the wavevector and energy dependence of the dipole correlations depends on
all exchange parameters.
It would be interesting to test the theory with data from larger scattering wavevectors, but that is beyond the scope of \cite{original-paper}.

\section{Negligibility of higher-order multipolar corrections to the scattering cross-section}
\label{sec:multipoles}

We now consider the neutron scattering cross section and show that the multipolar corrections discussed in \cite{lovesey-comment}
are negligibly small over the range of wavevectors we access.

The magnetic scattering cross section is related to the correlations of the Fourier transform of the magnetization operator $\hat{{\bf M}}({\bf q})$:
\begin{eqnarray}
\frac{d \sigma}{d \Omega} \propto \sum_{\alpha \beta} 
\left( \delta_{\alpha \beta} - \frac{q_{\alpha} q_{\beta}}{q^2} \right)
\langle \hat{M}_{\alpha} (-{\bf q}) \hat{M}_{\beta} ({\bf q}) \rangle
\label{eq:general_x_section}
\end{eqnarray}
where $\left( \delta_{\alpha \beta} - \frac{q_{\alpha} q_{\beta}}{q^2} \right)$ projects out the longitudinal component of $\hat{{\bf M}}$.
When the wavevector ${\bf q}$ is small enough that the scattering is not sensitive to atomic scale variations of the magnetisation, we can rewrite $\hat{{\bf M}}({\bf q})$ as:
\begin{eqnarray}
\hat{{\bf M}}({\bf q}) \approx f_{\rm dip} (q) {\bf m}({\bf q})
\label{eq:Mbasic}
\end{eqnarray}
where  ${\bf m}({\bf q})$ is the lattice Fourier transform of the ionic magnetic moments and $f_{\rm dip} (q)$ is a scalar form factor.
This is the dipole approximation \cite{marshall-lovesey, Rotter2009}.
In this approximation the form factor is:
\begin{equation}
f_{\rm dip} (q) = \langle j_0 (q) \rangle + \frac{(2-g_J)}{g_J}  \langle j_2 (q) \rangle
\label{eq:fdip}
\end{equation}
where $g_J$ is the Land{\'e} $g$-factor ($=\frac{6}{7}$ for Ce$^{3+}$) and $\langle j_n(q) \rangle$ are the expectation values of spherical Bessel
functions $J_n(q \ r)$ evaluated with respect to the radial electron density of the magnetic ions.

This leads to the following result for the scattering cross-section, from which Eq. (F1) of [\onlinecite{original-paper}] can be
derived (see Section \ref{sec:pseudospins}):
\begin{eqnarray}
\frac{d \sigma}{d \Omega} \propto |f_{\rm dip}(q)|^2 \sum_{\alpha \beta} 
\left( \delta_{\alpha \beta} - \frac{q_{\alpha} q_{\beta}}{q^2} \right)
\langle m_{\alpha} (-{\bf q}) m_{\beta} ({\bf q}) \rangle
\label{eq:dipole_x_section}
\end{eqnarray}

All of our calculations of neutron scattering responses in [\onlinecite{original-paper}] employ the dipole approximation
with the form factor being defined by Eq. (\ref{eq:fdip}). We employed analytical approximations for $\langle j_n(q) \rangle$ 
given in [\onlinecite{Lisher1971}].

To go beyond the dipole approximation, requires a more sophisticated treatment of $\hat{{\bf M}}({\bf q})$.
To achieve this, one can represent the matrix elements of $\hat{{\bf M}}({\bf q})$ between different angular momentum states
$|J, m_J\rangle$ as \cite{marshall-lovesey, Rotter2009}:
\begin{widetext}
\begin{eqnarray}
&&\langle J, m_J | \hat{M}_x | J, m_J' \rangle = 
-\sqrt{4 \pi} \mu_B \sum_{K=1,3,5...}  \frac{Z(K, q)}{K} P(K, m_J', m_J)
\big[ Y_{K-1, m_J'-m_J+1} (\tilde{\bf{q}}) \sqrt{(K+m_J-m_J')(K+m_J-m_J'-1)} \nonumber \\
&& \qquad \qquad \qquad \qquad -  Y_{K-1, m_J'-m_J-1} (\tilde{\bf{q}})  \sqrt{(K-m_J+m_J')(K-m_J+m_J'-1)}\big]
\nonumber \\
&&\langle J, m_J | \hat{M}_y | J, m_J' \rangle = i  \sqrt{4 \pi} \mu_B \sum_{K=1,3,5...} \frac{Z(K, q)}{K} P(K, m_J', m_J)
\big[ Y_{K-1, m_J'-m_J+1}(\tilde{\bf{q}})  \sqrt{(K+m_J-m_J')(K+m_J-m_J'-1)} \nonumber \\
&& \qquad \qquad \qquad \qquad  +  Y_{K-1, m_J'-m_J-1} (\tilde{\bf{q}}) \sqrt{(K-m_J+m_J')(K-m_J+m_J'-1)}\big]
\nonumber \\
&&\langle J, m_J | \hat{M}_z | J, m_J' \rangle =  -\sqrt{4 \pi} \mu_B \sum_{K=1,3,5...}  \frac{Z(K, q)}{K} P(K, m_J', m_J) Y_{K-1, m_J'-m_J} (\tilde{\bf{q}}) 
\sqrt{(K-m_J+m_J')(K+m_J-m_J')}
\label{eq:M_complicated}
\end{eqnarray}
\end{widetext}
where $Y_{nl}(\tilde{\bf{q}}) $ are spherical harmonic functions of the direction of ${\bf q}$, $Y_{nl}=0$ for $l>n$ and
\begin{eqnarray}
P(K, m_J, m_J')=(-1)^{J-m_J'}
\frac{\begin{pmatrix}
K&J&J\\
m_J-m_J'&m_J'&-m_J
\end{pmatrix}}{\begin{pmatrix}
K&J&J\\
0&J&-J
\end{pmatrix}}
\end{eqnarray}
with $\begin{pmatrix}
M_1&M_2&M_3\\
m_1 &m_2&m_3
\end{pmatrix}$
being Wigner $3j$-symbols.
The functions $Z(K, q)$ are linear combinations of the expectation values of
spherical Bessel functions $\langle j_n(q) \rangle$:
\begin{eqnarray}
Z(K, q) = c_{K-1} \langle j_{K-1} (q) \rangle + c_{K+1} \langle j_{K+1} (q) \rangle 
\end{eqnarray}
with the coefficients $c_{K}$ for rare-earth ions being given in Table 6.1 of [\onlinecite{marshall-lovesey}].

Considering only the $K=1$ term of Eq. (\ref{eq:M_complicated}) results in an equation of the same form
as the dipole approximation [Eq. (\ref{eq:Mbasic})], with a slightly different form factor $f'(q)$.
Using the coefficients $c_{K}$ given for Ce$^{3+}$ in  [\onlinecite{marshall-lovesey}], we obtain
\begin{eqnarray}
&&\hat{{\bf M}}({\bf q}) \approx f'(q) {\bf m}({\bf q}) \\
&& f'(q) = \langle j_0 (q) \rangle + \frac{8}{5} \langle j_2 (q) \rangle.
\end{eqnarray}
cf. $f_{\rm dip} (q)$ [Eq. (\ref{eq:fdip})] which differs in that the coefficient in front of $\langle j_2 (q) \rangle$ 
is $\frac{(2-g_J)}{g_J}=\frac{4}{3}$ for Ce$^{3+}$.

The terms involving higher values of $K$ contribute higher order multipole scattering terms, like those discussed
in \cite{lovesey-comment}. 
The $K=3$ and $K=5$ terms come with associated form factors:
\begin{eqnarray}
&&h(q)=\langle j_2 (q) \rangle + \frac{10}{3} \langle j_4 (q) \rangle \\
&&g(q)= \langle j_4 (q) \rangle + 12 \langle j_6 (q) \rangle
\end{eqnarray}
as found in [\onlinecite{lovesey-comment}].

Thus, the essential requirement for the validity of the dipole approximation is that $\langle j_0 (q) \rangle$ dominates
the other functions $\langle j_{n=2,4,6}(q) \rangle$ over the relevant range of $q$.
As long as this is the case neither the difference between $f'(q)$ and $f_{\rm dip}(q)$ nor the contributions of the $K>1$
terms will be important.

In Fig. \ref{fig:formfactors} we plot the squares of the form factors $f_{\rm dip}(q), f'(q), h(q), g(q)$, since the form factors ultimately
appear quadratically in the scattering intensity.
Over the range $q \in [0, 2.5] {\rm \AA}^{-1}$, which is the relevant range for our experiments, $f_{\rm dip}(q)^2$ agrees closely with $f'(q)^2$ and the higher-order contributions $h(q)^2$ and $g(q)^2$ are negligible.
This demonstrates the validity of our use of the dipole approximation in making theoretical calculations to compare with neutron scattering. 

\section{Use of pseudospins in describing physical properties}
\label{sec:pseudospins}

One of the foremost complaints in [\onlinecite{lovesey-comment}] is that we used pseudospin-$1/2$ operators to
describe the scattering, which the author argues have no physical relevance.
In this section, we will explain the use of these operators and demonstrate that they are appropriate for the purposes
for which we used them.
We note that the use of such operators to describe physical properties of 
rare-earth magnets is a well-established method in the literature, with a history of success, including
in the description of scattering experiments \cite{Ross2011, Savary2012_er, Benton2016, Li2015, rau16, Wen2017, RauReview2019, sarkis20, Sibille2020, Dun2020, samartzis22}.

As mentioned above, the crystal field interactions in Ce$_2$Zr$_2$O$_7$ separate a Kramers doublet from the rest of
the $J=5/2$ multiplet, by a gap $\Delta_{\rm CEF}\approx 56 {\rm meV}$ \cite{Gao2019, Gaudet2019}.
This should be compared to the temperature scales of our experiment $T\lesssim 10$ {\rm K} $\implies k_B T \lesssim 0.86 {\rm meV}$,
the estimated interaction energies $J_{\tilde{\alpha}} < 0.07$ {\rm meV}
and to the maximum energy transfers measured in our inelastic neutron scattering experiments $\omega \lesssim 0.4 \ {\rm meV}$.
Given these comparisons, we can be confident that the higher CEF levels are not significantly thermally populated in our experiment,
interactions do not appreciably mix higher CEF levels into the ground state and our inelastic experiments do not measure transitions
out of the low energy Kramers doublet.
Thus, all of the properties we measure in the experiment should be understood in terms of operators acting in the space of states
defined by the low energy Kramers doublet.
This is precisely the space in which the pseudospins act.

A basis for the low energy doublet is given by the $|J=5/2, m_J=\pm3/2\rangle$ states of the Ce$^{3+}$ ions, with the quantization axis ${\bf {\hat z}}$
being chosen to correspond with the local axis of $C_3$ rotational symmetry.
We can define a set of pseudospins in this basis $S^{\alpha}=\frac{1}{2}\sigma^{\alpha}$, with $\sigma^{\alpha}$ being the Pauli
matrices.
When combined with $2\times2$ identity matrix, these matrices provide a complete basis of hermitian $2\times2$ matrices.
This means that all physical processes which involve only the low energy states of the doublet can be represented as combinations
of these pseudospin operators.
Due to the temperature and energy scales of our experiment, this comprises all of the processes to which we are sensitive.
This is why the pseudospin operators are the appropriate tool for calculations.

Once projected into this low energy basis, all physical properties can be represented in the pseudospin language.
For example, defining $P$ to be the projector which projects into the low energy space, we see that the magnetic moment 
operator is:
\begin{eqnarray}
P {\bf m} P = g_J \mu_B P {\bf J} P = \mu_B \begin{pmatrix}
0\\
0\\
g_z S^z
\end{pmatrix}
\label{eq:moment}
\end{eqnarray}
where $g_z =18/7 \approx 2.57$.

Not only the magnetic moment, but higher order multipoles can also be represented in terms of pseudospins.
For example, we consider the octupolar operator $\overline{J_x J_x J_y}-J_y^3$ (where $\overline{abc}$ represents symmetrization over 
operator orderings). Projecting this into the low energy doublet gives
\begin{eqnarray}
P(\overline{J_x J_x J_y}-J_y^3)P=24 S^y.
\end{eqnarray}

All other higher-order multipoles,  can also be represented in this way once projected into the low energy doublet.
Thus, the pseudospin operators have simple and direct relations to physically measurable quantities.

We conclude this section of the reply by providing the derivation of Eq. (F1) of [\onlinecite{original-paper}]
which gives the neutron scattering intensity in terms of the correlations of pseudospins, within the dipole
approximation [Eq. (\ref{eq:dipole_x_section})].

The pseudospin basis used in [\onlinecite{original-paper}] is related to the one defined above, by a rotation around the
$y$-axis of pseudospin space by an angle $\theta$. This coordinate frame is chosen to remove a coupling between the $x$ and $z$
pseudospin components from the Hamiltonian (see [\onlinecite{Huang2014}, \onlinecite{Benton2016}]). 
The operators in the new coordinate frame $\tilde{x}, \tilde{y}, \tilde{z}$ are related to the original one $x,y,z$ as follows:
\begin{eqnarray}
&& S^x_i = \cos(\theta) S^{\tilde{x}}_i - \sin(\theta) S^{\tilde{z}}_i \\
&&S^y_i =  S^{\tilde{y}}_i \\
&&S^z_i =  \cos(\theta) S^{\tilde{z}}_i + \sin(\theta) S^{\tilde{x}}_i.
\end{eqnarray}

Combining this with Eq. (\ref{eq:moment}), we obtain the magnetic moment on each site as:
\begin{eqnarray}
{\bf m}_i = g_z \mu_B (\cos(\theta) S^{\tilde{z}}_i + \sin(\theta) S^{\tilde{x}}_i) \hat{{\bf z}}_i
\label{eq:moment_new}
\end{eqnarray}
where $\hat{{\bf z}}_i$ is the local axis of $C_3$ symmetry on site $i$.
There are four distinct sites in the unit cell of the pyrochlore lattice, defining four sublattices, 
with different anisotropy axes $\hat{{\bf z}}_i$.

We define a lattice Fourier transformed operator ${\bf m}_j({\bf q})$ as the Fourier transform over the sublattice $j$,
with $j=0,1,2,3$ being a sublattice index. Then ${\bf m}({\bf q})=\sum_j {\bf m}_j({\bf q})$.
Inserting this into Eq. (\ref{eq:dipole_x_section}) we obtain
\begin{widetext}
\begin{eqnarray}
S({\bf q})= g_z^2 \mu_B^2 f_{\rm dip}({\bf q})^2 \sum_{i,j,=0,1,2,3} \left( \hat{\bf z}_i \cdot \hat{\bf z}_j - \frac{(\hat{\bf z}_i  \cdot {\bf q})(\hat{\bf z}_j  \cdot {\bf q})}{q^2} \right) \langle  (\cos(\theta) S^{\tilde z}_i (-{\bf q})+ \sin(\theta) S^{\tilde x}_i(-{\bf q})) (\cos(\theta) S^{\tilde z}_j ({\bf q})+ \sin(\theta) S^{\tilde x}_j({\bf q})) \rangle \nonumber \\
\label{eq:intensity}
\end{eqnarray} 
Since we compare only relative, rather than absolute intensities with experiment we drop the multiplicative constant
$g_z^2\mu_B^2$. Finally we note that the symmetries of $XYZ$ Hamiltonian cause the cross correlators $\langle S^{\tilde z}(-{\bf q}) S^{\tilde x}({\bf q}) \rangle$ to vanish. 
Applying these points to Eq. (\ref{eq:intensity}) we obtain Eq. (F1) of [\onlinecite{original-paper}]:
\begin{eqnarray}
S({\bf q})= f_{\rm dip}({\bf q})^2 \sum_{i,j,=0,1,2,3} \left( \hat{\bf z}_i \cdot \hat{\bf z}_j - \frac{(\hat{\bf z}_i  \cdot {\bf q})(\hat{\bf z}_j  \cdot {\bf q})}{q^2} \right) 
\left[ \cos(\theta)^2 \langle   S^{\tilde z}_i (-{\bf q}) S^{\tilde z}_j ({\bf q}) \rangle + \sin(\theta)^2 \langle  
 S^{\tilde x}_i (-{\bf q})  S^{\tilde x}_j({\bf q})) \rangle \right] \nonumber \\
\label{eq:intensity-F1}
\end{eqnarray}
\end{widetext}

Eqs. (E1) and (E2) of [\onlinecite{original-paper}] follow similarly, with consideration of the neutron polarization.

\section{Inconsistency of the comment's calculations with experiment}
\label{sec:inconsistency}

Finally, we comment briefly on the calculations of the scattering amplitude in [\onlinecite{lovesey-comment}].
In the comment, the expectation value of the scattering operator ${\bf Q}$ (proportional to $\hat{{\bf M}}$ as given in Eq. (\ref{eq:M_complicated})) is calculated, assuming two different ground states.

In both cases the ground state of the many body system is assumed to be approximated as a product of single
site wavefunctions. 
The two single-site wavefunctions proposed are:
\begin{eqnarray}
&&|u \rangle= a |J=5/2, m_J=3/2 \rangle +  b |J=5/2, m_J=-3/2 \rangle \nonumber \\
\\
&&| g \rangle= \frac{1}{\sqrt{2}} \big( (a + ib)  |J=5/2, m_J=3/2 \rangle \nonumber \\
&&\qquad  + (b-ia) |J=5/2, m_J=-3/2 \rangle \big)
\end{eqnarray}
with $a$ and $b$ being unknown real coefficients and $a^2+b^2=1$.

It is important to note here that both $|u\rangle$ and $|g\rangle$ break time reversal symmetry.
This can be seen in the case of $|u\rangle$ by the generally finite expectation value of  $J_z$
\begin{eqnarray}
\langle u | J_z |u\rangle= \frac{3}{2} (a^2 - b^2 )
\end{eqnarray}
implying magnetic order. As noted in [\onlinecite{lovesey-comment}] this also breaks some lattice symmetries.
In fact, this would correspond with all-in-all-out antiferromagnetic order, which we do not observe.

The state $|g\rangle$ preserves all lattice symmetries, but nevertheless possesses a finite value of the
octupolar order parameter
\begin{eqnarray}
\langle g| \left( \overline{J_x J_x J_y} - J_y^3\right) | g \rangle =12
\end{eqnarray}
which breaks time reversal symmetry. This corresponds with the octupole ordered state on the phase diagram.

Since both these ground states break symmetry, reaching them from the high temperature paramagnet would
require a thermodynamic phase transition. No such phase transition is observed in either our heat capacity or
susceptibility measurements.
Muon spin relaxation measurements \cite{Gao2019}, also find no 
evidence of time reversal symmetry breaking.
Thus the assumptions behind the calculation in [\onlinecite{lovesey-comment}]
are inconsistent with experiment.

In the absence of time reversal symmetry breaking, the expectation value $\langle {\bf Q} \rangle$ vanishes
and the correlator  $\langle {\bf Q}  \cdot {\bf Q}\rangle$ cannot be approximated as $\langle {\bf Q} \rangle \cdot \langle {\bf Q} \rangle$
as suggested in [\onlinecite{lovesey-comment}].

\section{Summary}
\label{sec:summary}
 
 To summarize, the arguments advanced in [\onlinecite{lovesey-comment}] do not warrant any revision to
our analysis of the neutron scattering data in [\onlinecite{original-paper}].
 The multipolar corrections pointed out in  [\onlinecite{lovesey-comment}] are negligible  at the wavevectors
 we study, and the pseudospin formalism we use is appropriate for the temperature and energy range
 of our experiments.
 Further, the calculations of the scattering matrix elements in [\onlinecite{lovesey-comment}]  are themselves not
 consistent with experiment since they assume time-reversal symmetry breaking order,  requiring a thermodynamic
 phase transition on cooling from the high temperature paramagnet, which we do not observe.

\bibliography{reply_lovesey.bib}

\begin{thebibliography}{21}%
\makeatletter
\providecommand \@ifxundefined [1]{%
 \@ifx{#1\undefined}
}%
\providecommand \@ifnum [1]{%
 \ifnum #1\expandafter \@firstoftwo
 \else \expandafter \@secondoftwo
 \fi
}%
\providecommand \@ifx [1]{%
 \ifx #1\expandafter \@firstoftwo
 \else \expandafter \@secondoftwo
 \fi
}%
\providecommand \natexlab [1]{#1}%
\providecommand \enquote  [1]{``#1''}%
\providecommand \bibnamefont  [1]{#1}%
\providecommand \bibfnamefont [1]{#1}%
\providecommand \citenamefont [1]{#1}%
\providecommand \href@noop [0]{\@secondoftwo}%
\providecommand \href [0]{\begingroup \@sanitize@url \@href}%
\providecommand \@href[1]{\@@startlink{#1}\@@href}%
\providecommand \@@href[1]{\endgroup#1\@@endlink}%
\providecommand \@sanitize@url [0]{\catcode `\\12\catcode `\$12\catcode
  `\&12\catcode `\#12\catcode `\^12\catcode `\_12\catcode `\%12\relax}%
\providecommand \@@startlink[1]{}%
\providecommand \@@endlink[0]{}%
\providecommand \url  [0]{\begingroup\@sanitize@url \@url }%
\providecommand \@url [1]{\endgroup\@href {#1}{\urlprefix }}%
\providecommand \urlprefix  [0]{URL }%
\providecommand \Eprint [0]{\href }%
\providecommand \doibase [0]{https://doi.org/}%
\providecommand \selectlanguage [0]{\@gobble}%
\providecommand \bibinfo  [0]{\@secondoftwo}%
\providecommand \bibfield  [0]{\@secondoftwo}%
\providecommand \translation [1]{[#1]}%
\providecommand \BibitemOpen [0]{}%
\providecommand \bibitemStop [0]{}%
\providecommand \bibitemNoStop [0]{.\EOS\space}%
\providecommand \EOS [0]{\spacefactor3000\relax}%
\providecommand \BibitemShut  [1]{\csname bibitem#1\endcsname}%
\let\auto@bib@innerbib\@empty
\bibitem [{\citenamefont {Lovesey}()}]{lovesey-comment}%
  \BibitemOpen
  \bibfield  {author} {\bibinfo {author} {\bibfnamefont {S.~W.}\ \bibnamefont
  {Lovesey}},\ }\bibfield  {title} {\bibinfo {title} {Comment on "case for a
  u(1)pi quantum spin liquid ground state in the dipole-octupole pyrochlore
  ce2zr2o7" by e. m. smith et al., phys. rev. x 12, 021015 (2022)},\ }\href
  {https://arxiv.org/abs/2209.03235} {\bibinfo  {journal} {arXiv:2209.03235}\
  }\BibitemShut {NoStop}%
\bibitem [{\citenamefont {Smith}\ \emph {et~al.}(2022)\citenamefont {Smith},
  \citenamefont {Benton}, \citenamefont {Yahne}, \citenamefont {Placke},
  \citenamefont {Sch\"afer}, \citenamefont {Gaudet}, \citenamefont {Dudemaine},
  \citenamefont {Fitterman}, \citenamefont {Beare}, \citenamefont {Wildes},
  \citenamefont {Bhattacharya}, \citenamefont {DeLazzer}, \citenamefont
  {Buhariwalla}, \citenamefont {Butch}, \citenamefont {Movshovich},
  \citenamefont {Garrett}, \citenamefont {Marjerrison}, \citenamefont {Clancy},
  \citenamefont {Kermarrec}, \citenamefont {Luke}, \citenamefont {Bianchi},
  \citenamefont {Ross},\ and\ \citenamefont {Gaulin}}]{original-paper}%
  \BibitemOpen
\bibfield  {journal} {  }\bibfield  {author} {\bibinfo {author} {\bibfnamefont
  {E.~M.}\ \bibnamefont {Smith}}, \bibinfo {author} {\bibfnamefont
  {O.}~\bibnamefont {Benton}}, \bibinfo {author} {\bibfnamefont {D.~R.}\
  \bibnamefont {Yahne}}, \bibinfo {author} {\bibfnamefont {B.}~\bibnamefont
  {Placke}}, \bibinfo {author} {\bibfnamefont {R.}~\bibnamefont {Sch\"afer}},
  \bibinfo {author} {\bibfnamefont {J.}~\bibnamefont {Gaudet}}, \bibinfo
  {author} {\bibfnamefont {J.}~\bibnamefont {Dudemaine}}, \bibinfo {author}
  {\bibfnamefont {A.}~\bibnamefont {Fitterman}}, \bibinfo {author}
  {\bibfnamefont {J.}~\bibnamefont {Beare}}, \bibinfo {author} {\bibfnamefont
  {A.~R.}\ \bibnamefont {Wildes}}, \bibinfo {author} {\bibfnamefont
  {S.}~\bibnamefont {Bhattacharya}}, \bibinfo {author} {\bibfnamefont
  {T.}~\bibnamefont {DeLazzer}}, \bibinfo {author} {\bibfnamefont {C.~R.~C.}\
  \bibnamefont {Buhariwalla}}, \bibinfo {author} {\bibfnamefont {N.~P.}\
  \bibnamefont {Butch}}, \bibinfo {author} {\bibfnamefont {R.}~\bibnamefont
  {Movshovich}}, \bibinfo {author} {\bibfnamefont {J.~D.}\ \bibnamefont
  {Garrett}}, \bibinfo {author} {\bibfnamefont {C.~A.}\ \bibnamefont
  {Marjerrison}}, \bibinfo {author} {\bibfnamefont {J.~P.}\ \bibnamefont
  {Clancy}}, \bibinfo {author} {\bibfnamefont {E.}~\bibnamefont {Kermarrec}},
  \bibinfo {author} {\bibfnamefont {G.~M.}\ \bibnamefont {Luke}}, \bibinfo
  {author} {\bibfnamefont {A.~D.}\ \bibnamefont {Bianchi}}, \bibinfo {author}
  {\bibfnamefont {K.~A.}\ \bibnamefont {Ross}},\ and\ \bibinfo {author}
  {\bibfnamefont {B.~D.}\ \bibnamefont {Gaulin}},\ }\bibfield  {title}
  {\bibinfo {title} {Case for a ${\mathrm{u}(1)}_{\ensuremath{\pi}}$ quantum
  spin liquid ground state in the dipole-octupole pyrochlore
  ${\mathrm{ce}}_{2}{\mathrm{zr}}_{2}{\mathrm{o}}_{7}$},\ }\href
  {https://doi.org/10.1103/PhysRevX.12.021015} {\bibfield  {journal} {\bibinfo
  {journal} {Phys. Rev. X}\ }\textbf {\bibinfo {volume} {12}},\ \bibinfo
  {pages} {021015} (\bibinfo {year} {2022})}\BibitemShut {NoStop}%
\bibitem [{\citenamefont {Ross}\ \emph {et~al.}(2011)\citenamefont {Ross},
  \citenamefont {Savary}, \citenamefont {Gaulin},\ and\ \citenamefont
  {Balents}}]{Ross2011}%
  \BibitemOpen
  \bibfield  {author} {\bibinfo {author} {\bibfnamefont {K.~A.}\ \bibnamefont
  {Ross}}, \bibinfo {author} {\bibfnamefont {L.}~\bibnamefont {Savary}},
  \bibinfo {author} {\bibfnamefont {B.~D.}\ \bibnamefont {Gaulin}},\ and\
  \bibinfo {author} {\bibfnamefont {L.}~\bibnamefont {Balents}},\ }\bibfield
  {title} {\bibinfo {title} {Quantum excitations in quantum spin ice},\ }\href
  {https://doi.org/10.1103/PhysRevX.1.021002} {\bibfield  {journal} {\bibinfo
  {journal} {Phys. Rev. X}\ }\textbf {\bibinfo {volume} {1}},\ \bibinfo {pages}
  {021002} (\bibinfo {year} {2011})}\BibitemShut {NoStop}%
\bibitem [{\citenamefont {Savary}\ \emph {et~al.}(2012)\citenamefont {Savary},
  \citenamefont {Ross}, \citenamefont {Gaulin}, \citenamefont {Ruff},\ and\
  \citenamefont {Balents}}]{Savary2012_er}%
  \BibitemOpen
  \bibfield  {author} {\bibinfo {author} {\bibfnamefont {L.}~\bibnamefont
  {Savary}}, \bibinfo {author} {\bibfnamefont {K.~A.}\ \bibnamefont {Ross}},
  \bibinfo {author} {\bibfnamefont {B.~D.}\ \bibnamefont {Gaulin}}, \bibinfo
  {author} {\bibfnamefont {J.~P.~C.}\ \bibnamefont {Ruff}},\ and\ \bibinfo
  {author} {\bibfnamefont {L.}~\bibnamefont {Balents}},\ }\bibfield  {title}
  {\bibinfo {title} {Order by quantum disorder in
  ${\mathrm{er}}_{2}{\mathrm{ti}}_{2}{\mathbf{o}}_{7}$},\ }\href
  {https://doi.org/10.1103/PhysRevLett.109.167201} {\bibfield  {journal}
  {\bibinfo  {journal} {Phys. Rev. Lett.}\ }\textbf {\bibinfo {volume} {109}},\
  \bibinfo {pages} {167201} (\bibinfo {year} {2012})}\BibitemShut {NoStop}%
\bibitem [{\citenamefont {Benton}(2016)}]{Benton2016}%
  \BibitemOpen
  \bibfield  {author} {\bibinfo {author} {\bibfnamefont {O.}~\bibnamefont
  {Benton}},\ }\bibfield  {title} {\bibinfo {title} {Quantum origins of moment
  fragmentation in ${\mathrm{nd}}_{2}{\mathrm{zr}}_{2}{\mathrm{o}}_{7}$},\
  }\href {https://doi.org/10.1103/PhysRevB.94.104430} {\bibfield  {journal}
  {\bibinfo  {journal} {Phys. Rev. B}\ }\textbf {\bibinfo {volume} {94}},\
  \bibinfo {pages} {104430} (\bibinfo {year} {2016})}\BibitemShut {NoStop}%
\bibitem [{\citenamefont {Li}\ \emph {et~al.}(2015)\citenamefont {Li},
  \citenamefont {Chen}, \citenamefont {Tong}, \citenamefont {Pi}, \citenamefont
  {Liu}, \citenamefont {Yang}, \citenamefont {Wang},\ and\ \citenamefont
  {Zhang}}]{Li2015}%
  \BibitemOpen
  \bibfield  {author} {\bibinfo {author} {\bibfnamefont {Y.}~\bibnamefont
  {Li}}, \bibinfo {author} {\bibfnamefont {G.}~\bibnamefont {Chen}}, \bibinfo
  {author} {\bibfnamefont {W.}~\bibnamefont {Tong}}, \bibinfo {author}
  {\bibfnamefont {L.}~\bibnamefont {Pi}}, \bibinfo {author} {\bibfnamefont
  {J.}~\bibnamefont {Liu}}, \bibinfo {author} {\bibfnamefont {Z.}~\bibnamefont
  {Yang}}, \bibinfo {author} {\bibfnamefont {X.}~\bibnamefont {Wang}},\ and\
  \bibinfo {author} {\bibfnamefont {Q.}~\bibnamefont {Zhang}},\ }\bibfield
  {title} {\bibinfo {title} {Rare-earth triangular lattice spin liquid: A
  single-crystal study of ${\mathrm{ybmggao}}_{4}$},\ }\href
  {https://doi.org/10.1103/PhysRevLett.115.167203} {\bibfield  {journal}
  {\bibinfo  {journal} {Phys. Rev. Lett.}\ }\textbf {\bibinfo {volume} {115}},\
  \bibinfo {pages} {167203} (\bibinfo {year} {2015})}\BibitemShut {NoStop}%
\bibitem [{\citenamefont {Rau}\ \emph {et~al.}(2016)\citenamefont {Rau},
  \citenamefont {Wu}, \citenamefont {May}, \citenamefont {Poudel},
  \citenamefont {Winn}, \citenamefont {Garlea}, \citenamefont {Huq},
  \citenamefont {Whitfield}, \citenamefont {Taylor}, \citenamefont {Lumsden},
  \citenamefont {Gingras},\ and\ \citenamefont {Christianson}}]{rau16}%
  \BibitemOpen
  \bibfield  {author} {\bibinfo {author} {\bibfnamefont {J.~G.}\ \bibnamefont
  {Rau}}, \bibinfo {author} {\bibfnamefont {L.~S.}\ \bibnamefont {Wu}},
  \bibinfo {author} {\bibfnamefont {A.~F.}\ \bibnamefont {May}}, \bibinfo
  {author} {\bibfnamefont {L.}~\bibnamefont {Poudel}}, \bibinfo {author}
  {\bibfnamefont {B.}~\bibnamefont {Winn}}, \bibinfo {author} {\bibfnamefont
  {V.~O.}\ \bibnamefont {Garlea}}, \bibinfo {author} {\bibfnamefont
  {A.}~\bibnamefont {Huq}}, \bibinfo {author} {\bibfnamefont {P.}~\bibnamefont
  {Whitfield}}, \bibinfo {author} {\bibfnamefont {A.~E.}\ \bibnamefont
  {Taylor}}, \bibinfo {author} {\bibfnamefont {M.~D.}\ \bibnamefont {Lumsden}},
  \bibinfo {author} {\bibfnamefont {M.~J.~P.}\ \bibnamefont {Gingras}},\ and\
  \bibinfo {author} {\bibfnamefont {A.~D.}\ \bibnamefont {Christianson}},\
  }\bibfield  {title} {\bibinfo {title} {Anisotropic exchange within decoupled
  tetrahedra in the quantum breathing pyrochlore
  ${\mathrm{ba}}_{3}{\mathrm{yb}}_{2}{\mathrm{zn}}_{5}{\mathrm{o}}_{11}$},\
  }\href {https://doi.org/10.1103/PhysRevLett.116.257204} {\bibfield  {journal}
  {\bibinfo  {journal} {Phys. Rev. Lett.}\ }\textbf {\bibinfo {volume} {116}},\
  \bibinfo {pages} {257204} (\bibinfo {year} {2016})}\BibitemShut {NoStop}%
\bibitem [{\citenamefont {Wen}\ \emph {et~al.}(2017)\citenamefont {Wen},
  \citenamefont {Koohpayeh}, \citenamefont {Ross}, \citenamefont {Trump},
  \citenamefont {McQueen}, \citenamefont {Kimura}, \citenamefont {Nakatsuji},
  \citenamefont {Qiu}, \citenamefont {Pajerowski}, \citenamefont {Copley},\
  and\ \citenamefont {Broholm}}]{Wen2017}%
  \BibitemOpen
  \bibfield  {author} {\bibinfo {author} {\bibfnamefont {J.-J.}\ \bibnamefont
  {Wen}}, \bibinfo {author} {\bibfnamefont {S.~M.}\ \bibnamefont {Koohpayeh}},
  \bibinfo {author} {\bibfnamefont {K.~A.}\ \bibnamefont {Ross}}, \bibinfo
  {author} {\bibfnamefont {B.~A.}\ \bibnamefont {Trump}}, \bibinfo {author}
  {\bibfnamefont {T.~M.}\ \bibnamefont {McQueen}}, \bibinfo {author}
  {\bibfnamefont {K.}~\bibnamefont {Kimura}}, \bibinfo {author} {\bibfnamefont
  {S.}~\bibnamefont {Nakatsuji}}, \bibinfo {author} {\bibfnamefont
  {Y.}~\bibnamefont {Qiu}}, \bibinfo {author} {\bibfnamefont {D.~M.}\
  \bibnamefont {Pajerowski}}, \bibinfo {author} {\bibfnamefont {J.~R.~D.}\
  \bibnamefont {Copley}},\ and\ \bibinfo {author} {\bibfnamefont {C.~L.}\
  \bibnamefont {Broholm}},\ }\bibfield  {title} {\bibinfo {title} {Disordered
  route to the coulomb quantum spin liquid: Random transverse fields on spin
  ice in ${\mathrm{pr}}_{2}{\mathrm{zr}}_{2}{\mathrm{o}}_{7}$},\ }\href
  {https://doi.org/10.1103/PhysRevLett.118.107206} {\bibfield  {journal}
  {\bibinfo  {journal} {Phys. Rev. Lett.}\ }\textbf {\bibinfo {volume} {118}},\
  \bibinfo {pages} {107206} (\bibinfo {year} {2017})}\BibitemShut {NoStop}%
\bibitem [{\citenamefont {Rau}\ and\ \citenamefont
  {Gingras}(2019)}]{RauReview2019}%
  \BibitemOpen
  \bibfield  {author} {\bibinfo {author} {\bibfnamefont {J.~G.}\ \bibnamefont
  {Rau}}\ and\ \bibinfo {author} {\bibfnamefont {M.~J.}\ \bibnamefont
  {Gingras}},\ }\bibfield  {title} {\bibinfo {title} {Frustrated quantum
  rare-earth pyrochlores},\ }\href
  {https://doi.org/10.1146/annurev-conmatphys-022317-110520} {\bibfield
  {journal} {\bibinfo  {journal} {Annu. Rev. Condens. Matter Phys}\ }\textbf
  {\bibinfo {volume} {10}},\ \bibinfo {pages} {357} (\bibinfo {year}
  {2019})}\BibitemShut {NoStop}%
\bibitem [{\citenamefont {Sarkis}\ \emph {et~al.}(2020)\citenamefont {Sarkis},
  \citenamefont {Rau}, \citenamefont {Sanjeewa}, \citenamefont {Powell},
  \citenamefont {Kolis}, \citenamefont {Marbey}, \citenamefont {Hill},
  \citenamefont {Rodriguez-Rivera}, \citenamefont {Nair}, \citenamefont
  {Yahne}, \citenamefont {S\"aubert}, \citenamefont {Gingras},\ and\
  \citenamefont {Ross}}]{sarkis20}%
  \BibitemOpen
  \bibfield  {author} {\bibinfo {author} {\bibfnamefont {C.~L.}\ \bibnamefont
  {Sarkis}}, \bibinfo {author} {\bibfnamefont {J.~G.}\ \bibnamefont {Rau}},
  \bibinfo {author} {\bibfnamefont {L.~D.}\ \bibnamefont {Sanjeewa}}, \bibinfo
  {author} {\bibfnamefont {M.}~\bibnamefont {Powell}}, \bibinfo {author}
  {\bibfnamefont {J.}~\bibnamefont {Kolis}}, \bibinfo {author} {\bibfnamefont
  {J.}~\bibnamefont {Marbey}}, \bibinfo {author} {\bibfnamefont
  {S.}~\bibnamefont {Hill}}, \bibinfo {author} {\bibfnamefont {J.~A.}\
  \bibnamefont {Rodriguez-Rivera}}, \bibinfo {author} {\bibfnamefont {H.~S.}\
  \bibnamefont {Nair}}, \bibinfo {author} {\bibfnamefont {D.~R.}\ \bibnamefont
  {Yahne}}, \bibinfo {author} {\bibfnamefont {S.}~\bibnamefont {S\"aubert}},
  \bibinfo {author} {\bibfnamefont {M.~J.~P.}\ \bibnamefont {Gingras}},\ and\
  \bibinfo {author} {\bibfnamefont {K.~A.}\ \bibnamefont {Ross}},\ }\bibfield
  {title} {\bibinfo {title} {Unravelling competing microscopic interactions at
  a phase boundary: A single-crystal study of the metastable antiferromagnetic
  pyrochlore ${\mathrm{yb}}_{2}{\mathrm{ge}}_{2}{\mathrm{o}}_{7}$},\ }\href
  {https://doi.org/10.1103/PhysRevB.102.134418} {\bibfield  {journal} {\bibinfo
   {journal} {Phys. Rev. B}\ }\textbf {\bibinfo {volume} {102}},\ \bibinfo
  {pages} {134418} (\bibinfo {year} {2020})}\BibitemShut {NoStop}%
\bibitem [{\citenamefont {Sibille}\ \emph {et~al.}(2020)\citenamefont
  {Sibille}, \citenamefont {Gauthier}, \citenamefont {Lhotel}, \citenamefont
  {Porée}, \citenamefont {Pomjakushin}, \citenamefont {Ewings}, \citenamefont
  {Perring}, \citenamefont {Ollivier}, \citenamefont {Wildes}, \citenamefont
  {Ritter}, \citenamefont {Hansen}, \citenamefont {Keen}, \citenamefont
  {Nilsen}, \citenamefont {Keller}, \citenamefont {Petit},\ and\ \citenamefont
  {Fennell}}]{Sibille2020}%
  \BibitemOpen
  \bibfield  {author} {\bibinfo {author} {\bibfnamefont {R.}~\bibnamefont
  {Sibille}}, \bibinfo {author} {\bibfnamefont {N.}~\bibnamefont {Gauthier}},
  \bibinfo {author} {\bibfnamefont {E.}~\bibnamefont {Lhotel}}, \bibinfo
  {author} {\bibfnamefont {V.}~\bibnamefont {Porée}}, \bibinfo {author}
  {\bibfnamefont {V.}~\bibnamefont {Pomjakushin}}, \bibinfo {author}
  {\bibfnamefont {R.}~\bibnamefont {Ewings}}, \bibinfo {author} {\bibfnamefont
  {T.}~\bibnamefont {Perring}}, \bibinfo {author} {\bibfnamefont
  {J.}~\bibnamefont {Ollivier}}, \bibinfo {author} {\bibfnamefont
  {A.}~\bibnamefont {Wildes}}, \bibinfo {author} {\bibfnamefont
  {C.}~\bibnamefont {Ritter}}, \bibinfo {author} {\bibfnamefont
  {T.}~\bibnamefont {Hansen}}, \bibinfo {author} {\bibfnamefont
  {D.}~\bibnamefont {Keen}}, \bibinfo {author} {\bibfnamefont {G.}~\bibnamefont
  {Nilsen}}, \bibinfo {author} {\bibfnamefont {L.}~\bibnamefont {Keller}},
  \bibinfo {author} {\bibfnamefont {S.}~\bibnamefont {Petit}},\ and\ \bibinfo
  {author} {\bibfnamefont {T.}~\bibnamefont {Fennell}},\ }\bibfield  {title}
  {\bibinfo {title} {A quantum liquid of magnetic octupoles on the pyrochlore
  lattice},\ }\href {https://www.nature.com/articles/s41567-020-0827-7}
  {\bibfield  {journal} {\bibinfo  {journal} {Nat. Phys.}\ }\textbf {\bibinfo
  {volume} {16}},\ \bibinfo {pages} {546–552} (\bibinfo {year}
  {2020})}\BibitemShut {NoStop}%
\bibitem [{\citenamefont {Dun}\ \emph {et~al.}(2020)\citenamefont {Dun},
  \citenamefont {Bai}, \citenamefont {Paddison}, \citenamefont {Hollingworth},
  \citenamefont {Butch}, \citenamefont {Cruz}, \citenamefont {Stone},
  \citenamefont {Hong}, \citenamefont {Demmel}, \citenamefont {Mourigal},\ and\
  \citenamefont {Zhou}}]{Dun2020}%
  \BibitemOpen
  \bibfield  {author} {\bibinfo {author} {\bibfnamefont {Z.}~\bibnamefont
  {Dun}}, \bibinfo {author} {\bibfnamefont {X.}~\bibnamefont {Bai}}, \bibinfo
  {author} {\bibfnamefont {J.~A.~M.}\ \bibnamefont {Paddison}}, \bibinfo
  {author} {\bibfnamefont {E.}~\bibnamefont {Hollingworth}}, \bibinfo {author}
  {\bibfnamefont {N.~P.}\ \bibnamefont {Butch}}, \bibinfo {author}
  {\bibfnamefont {C.~D.}\ \bibnamefont {Cruz}}, \bibinfo {author}
  {\bibfnamefont {M.~B.}\ \bibnamefont {Stone}}, \bibinfo {author}
  {\bibfnamefont {T.}~\bibnamefont {Hong}}, \bibinfo {author} {\bibfnamefont
  {F.}~\bibnamefont {Demmel}}, \bibinfo {author} {\bibfnamefont
  {M.}~\bibnamefont {Mourigal}},\ and\ \bibinfo {author} {\bibfnamefont
  {H.}~\bibnamefont {Zhou}},\ }\bibfield  {title} {\bibinfo {title} {Quantum
  versus classical spin fragmentation in dipolar kagome ice
  ${\mathrm{ho}}_{3}{\mathrm{mg}}_{2}{\mathrm{sb}}_{3}{\mathrm{o}}_{14}$},\
  }\href {https://doi.org/10.1103/PhysRevX.10.031069} {\bibfield  {journal}
  {\bibinfo  {journal} {Phys. Rev. X}\ }\textbf {\bibinfo {volume} {10}},\
  \bibinfo {pages} {031069} (\bibinfo {year} {2020})}\BibitemShut {NoStop}%
\bibitem [{\citenamefont {Samartzis}\ \emph {et~al.}(2022)\citenamefont
  {Samartzis}, \citenamefont {Xu}, \citenamefont {Anand}, \citenamefont
  {Islam}, \citenamefont {Ollivier}, \citenamefont {Su},\ and\ \citenamefont
  {Lake}}]{samartzis22}%
  \BibitemOpen
  \bibfield  {author} {\bibinfo {author} {\bibfnamefont {A.}~\bibnamefont
  {Samartzis}}, \bibinfo {author} {\bibfnamefont {J.}~\bibnamefont {Xu}},
  \bibinfo {author} {\bibfnamefont {V.~K.}\ \bibnamefont {Anand}}, \bibinfo
  {author} {\bibfnamefont {A.~T. M.~N.}\ \bibnamefont {Islam}}, \bibinfo
  {author} {\bibfnamefont {J.}~\bibnamefont {Ollivier}}, \bibinfo {author}
  {\bibfnamefont {Y.}~\bibnamefont {Su}},\ and\ \bibinfo {author}
  {\bibfnamefont {B.}~\bibnamefont {Lake}},\ }\bibfield  {title} {\bibinfo
  {title} {Pinch points and half-moons in dipolar-octupolar
  ${\mathrm{nd}}_{2}{\mathrm{hf}}_{2}{\mathrm{o}}_{7}$},\ }\href
  {https://doi.org/10.1103/PhysRevB.106.L100401} {\bibfield  {journal}
  {\bibinfo  {journal} {Phys. Rev. B}\ }\textbf {\bibinfo {volume} {106}},\
  \bibinfo {pages} {L100401} (\bibinfo {year} {2022})}\BibitemShut {NoStop}%
\bibitem [{\citenamefont {Gao}\ \emph {et~al.}(2019)\citenamefont {Gao},
  \citenamefont {Chen}, \citenamefont {Tam}, \citenamefont {Huang},
  \citenamefont {Sasmal}, \citenamefont {Adroja}, \citenamefont {Ye},
  \citenamefont {Cao}, \citenamefont {Sala}, \citenamefont {Stone},
  \citenamefont {Baines}, \citenamefont {Verezhak}, \citenamefont {Hu},
  \citenamefont {Chung}, \citenamefont {Xu}, \citenamefont {Cheong},
  \citenamefont {Nallaiyan}, \citenamefont {Spagna}, \citenamefont {Maple},
  \citenamefont {Nevidomskyy}, \citenamefont {Morosan}, \citenamefont {Chen},\
  and\ \citenamefont {Dai}}]{Gao2019}%
  \BibitemOpen
  \bibfield  {author} {\bibinfo {author} {\bibfnamefont {B.}~\bibnamefont
  {Gao}}, \bibinfo {author} {\bibfnamefont {T.}~\bibnamefont {Chen}}, \bibinfo
  {author} {\bibfnamefont {D.}~\bibnamefont {Tam}}, \bibinfo {author}
  {\bibfnamefont {C.-L.}\ \bibnamefont {Huang}}, \bibinfo {author}
  {\bibfnamefont {K.}~\bibnamefont {Sasmal}}, \bibinfo {author} {\bibfnamefont
  {D.}~\bibnamefont {Adroja}}, \bibinfo {author} {\bibfnamefont
  {F.}~\bibnamefont {Ye}}, \bibinfo {author} {\bibfnamefont {H.}~\bibnamefont
  {Cao}}, \bibinfo {author} {\bibfnamefont {G.}~\bibnamefont {Sala}}, \bibinfo
  {author} {\bibfnamefont {M.}~\bibnamefont {Stone}}, \bibinfo {author}
  {\bibfnamefont {C.}~\bibnamefont {Baines}}, \bibinfo {author} {\bibfnamefont
  {J.}~\bibnamefont {Verezhak}}, \bibinfo {author} {\bibfnamefont
  {H.}~\bibnamefont {Hu}}, \bibinfo {author} {\bibfnamefont {J.-H.}\
  \bibnamefont {Chung}}, \bibinfo {author} {\bibfnamefont {X.}~\bibnamefont
  {Xu}}, \bibinfo {author} {\bibfnamefont {S.-W.}\ \bibnamefont {Cheong}},
  \bibinfo {author} {\bibfnamefont {M.}~\bibnamefont {Nallaiyan}}, \bibinfo
  {author} {\bibfnamefont {S.}~\bibnamefont {Spagna}}, \bibinfo {author}
  {\bibfnamefont {M.}~\bibnamefont {Maple}}, \bibinfo {author} {\bibfnamefont
  {A.}~\bibnamefont {Nevidomskyy}}, \bibinfo {author} {\bibfnamefont
  {E.}~\bibnamefont {Morosan}}, \bibinfo {author} {\bibfnamefont
  {G.}~\bibnamefont {Chen}},\ and\ \bibinfo {author} {\bibfnamefont
  {P.}~\bibnamefont {Dai}},\ }\bibfield  {title} {\bibinfo {title}
  {Experimental signatures of a three-dimensional quantum spin liquid in
  effective spin-1/2 {Ce$_2$Zr$_2$O$_7$} pyrochlore},\ }\href
  {https://doi.org/10.1038/s41567-019-0577-6} {\bibfield  {journal} {\bibinfo
  {journal} {Nat. Phys.}\ }\textbf {\bibinfo {volume} {15}},\ \bibinfo {pages}
  {1052–1057} (\bibinfo {year} {2019})}\BibitemShut {NoStop}%
\bibitem [{\citenamefont {Gaudet}\ \emph {et~al.}(2019)\citenamefont {Gaudet},
  \citenamefont {Smith}, \citenamefont {Dudemaine}, \citenamefont {Beare},
  \citenamefont {Buhariwalla}, \citenamefont {Butch}, \citenamefont {Stone},
  \citenamefont {Kolesnikov}, \citenamefont {Xu}, \citenamefont {Yahne},
  \citenamefont {Ross}, \citenamefont {Marjerrison}, \citenamefont {Garrett},
  \citenamefont {Luke}, \citenamefont {Bianchi},\ and\ \citenamefont
  {Gaulin}}]{Gaudet2019}%
  \BibitemOpen
  \bibfield  {author} {\bibinfo {author} {\bibfnamefont {J.}~\bibnamefont
  {Gaudet}}, \bibinfo {author} {\bibfnamefont {E.~M.}\ \bibnamefont {Smith}},
  \bibinfo {author} {\bibfnamefont {J.}~\bibnamefont {Dudemaine}}, \bibinfo
  {author} {\bibfnamefont {J.}~\bibnamefont {Beare}}, \bibinfo {author}
  {\bibfnamefont {C.~R.~C.}\ \bibnamefont {Buhariwalla}}, \bibinfo {author}
  {\bibfnamefont {N.~P.}\ \bibnamefont {Butch}}, \bibinfo {author}
  {\bibfnamefont {M.~B.}\ \bibnamefont {Stone}}, \bibinfo {author}
  {\bibfnamefont {A.~I.}\ \bibnamefont {Kolesnikov}}, \bibinfo {author}
  {\bibfnamefont {G.}~\bibnamefont {Xu}}, \bibinfo {author} {\bibfnamefont
  {D.~R.}\ \bibnamefont {Yahne}}, \bibinfo {author} {\bibfnamefont {K.~A.}\
  \bibnamefont {Ross}}, \bibinfo {author} {\bibfnamefont {C.~A.}\ \bibnamefont
  {Marjerrison}}, \bibinfo {author} {\bibfnamefont {J.~D.}\ \bibnamefont
  {Garrett}}, \bibinfo {author} {\bibfnamefont {G.~M.}\ \bibnamefont {Luke}},
  \bibinfo {author} {\bibfnamefont {A.~D.}\ \bibnamefont {Bianchi}},\ and\
  \bibinfo {author} {\bibfnamefont {B.~D.}\ \bibnamefont {Gaulin}},\ }\bibfield
   {title} {\bibinfo {title} {Quantum spin ice dynamics in the dipole-octupole
  pyrochlore magnet {Ce$_2$Zr$_2$O$_7$}},\ }\href
  {https://doi.org/10.1103/PhysRevLett.122.187201} {\bibfield  {journal}
  {\bibinfo  {journal} {Phys. Rev. Lett.}\ }\textbf {\bibinfo {volume} {122}},\
  \bibinfo {pages} {187201} (\bibinfo {year} {2019})}\BibitemShut {NoStop}%
\bibitem [{\citenamefont {Huang}\ \emph {et~al.}(2014)\citenamefont {Huang},
  \citenamefont {Chen},\ and\ \citenamefont {Hermele}}]{Huang2014}%
  \BibitemOpen
  \bibfield  {author} {\bibinfo {author} {\bibfnamefont {Y.-P.}\ \bibnamefont
  {Huang}}, \bibinfo {author} {\bibfnamefont {G.}~\bibnamefont {Chen}},\ and\
  \bibinfo {author} {\bibfnamefont {M.}~\bibnamefont {Hermele}},\ }\bibfield
  {title} {\bibinfo {title} {Quantum spin ices and topological phases from
  dipolar-octupolar doublets on the pyrochlore lattice},\ }\href
  {https://doi.org/10.1103/PhysRevLett.112.167203} {\bibfield  {journal}
  {\bibinfo  {journal} {Phys. Rev. Lett.}\ }\textbf {\bibinfo {volume} {112}},\
  \bibinfo {pages} {167203} (\bibinfo {year} {2014})}\BibitemShut {NoStop}%
\bibitem [{\citenamefont {Benton}(2020)}]{Benton2020}%
  \BibitemOpen
  \bibfield  {author} {\bibinfo {author} {\bibfnamefont {O.}~\bibnamefont
  {Benton}},\ }\bibfield  {title} {\bibinfo {title} {Ground-state phase diagram
  of dipolar-octupolar pyrochlores},\ }\href
  {https://doi.org/10.1103/PhysRevB.102.104408} {\bibfield  {journal} {\bibinfo
   {journal} {Phys. Rev. B}\ }\textbf {\bibinfo {volume} {102}},\ \bibinfo
  {pages} {104408} (\bibinfo {year} {2020})}\BibitemShut {NoStop}%
\bibitem [{\citenamefont {Patri}\ \emph {et~al.}(2020)\citenamefont {Patri},
  \citenamefont {Hosoi},\ and\ \citenamefont {Kim}}]{Patri2020}%
  \BibitemOpen
  \bibfield  {author} {\bibinfo {author} {\bibfnamefont {A.~S.}\ \bibnamefont
  {Patri}}, \bibinfo {author} {\bibfnamefont {M.}~\bibnamefont {Hosoi}},\ and\
  \bibinfo {author} {\bibfnamefont {Y.~B.}\ \bibnamefont {Kim}},\ }\bibfield
  {title} {\bibinfo {title} {Distinguishing dipolar and octupolar quantum spin
  ices using contrasting magnetostriction signatures},\ }\href
  {https://doi.org/10.1103/PhysRevResearch.2.023253} {\bibfield  {journal}
  {\bibinfo  {journal} {Phys. Rev. Research}\ }\textbf {\bibinfo {volume}
  {2}},\ \bibinfo {pages} {023253} (\bibinfo {year} {2020})}\BibitemShut
  {NoStop}%
\bibitem [{\citenamefont {Lisher}\ and\ \citenamefont
  {Forsyth}(1971)}]{Lisher1971}%
  \BibitemOpen
  \bibfield  {author} {\bibinfo {author} {\bibfnamefont {E.~J.}\ \bibnamefont
  {Lisher}}\ and\ \bibinfo {author} {\bibfnamefont {J.~B.}\ \bibnamefont
  {Forsyth}},\ }\bibfield  {title} {\bibinfo {title} {{Analytic approximations
  to form factors}},\ }\href {https://doi.org/10.1107/S0567739471001219}
  {\bibfield  {journal} {\bibinfo  {journal} {Acta Crystallogr., Sect. A.}\
  }\textbf {\bibinfo {volume} {27}},\ \bibinfo {pages} {545} (\bibinfo {year}
  {1971})}\BibitemShut {NoStop}%
\bibitem [{\citenamefont {Marshall}\ and\ \citenamefont
  {Lovesey}(1971)}]{marshall-lovesey}%
  \BibitemOpen
  \bibfield  {author} {\bibinfo {author} {\bibfnamefont {W.}~\bibnamefont
  {Marshall}}\ and\ \bibinfo {author} {\bibfnamefont {S.~W.}\ \bibnamefont
  {Lovesey}},\ }\href@noop {} {\emph {\bibinfo {title} {Theory of Thermal
  Neutron Scattering}}}\ (\bibinfo  {publisher} {Oxford Universiy Press},\
  \bibinfo {year} {1971})\BibitemShut {NoStop}%
\bibitem [{\citenamefont {Rotter}\ and\ \citenamefont
  {Boothroyd}(2009)}]{Rotter2009}%
  \BibitemOpen
  \bibfield  {author} {\bibinfo {author} {\bibfnamefont {M.}~\bibnamefont
  {Rotter}}\ and\ \bibinfo {author} {\bibfnamefont {A.~T.}\ \bibnamefont
  {Boothroyd}},\ }\bibfield  {title} {\bibinfo {title} {Going beyond the dipole
  approximation to improve the refinement of magnetic structures by neutron
  diffraction},\ }\href {https://doi.org/10.1103/PhysRevB.79.140405} {\bibfield
   {journal} {\bibinfo  {journal} {Phys. Rev. B}\ }\textbf {\bibinfo {volume}
  {79}},\ \bibinfo {pages} {140405} (\bibinfo {year} {2009})}\BibitemShut
  {NoStop}%
\end{thebibliography}%

\end{document}